# Reinforcement Learning from Human Feedback: Whose Culture, Whose Values, Whose Perspectives?

Kristian González Barman[†a] , Simon Lohse[†♠ b] & Henk de Regt[b]

[†] Authors contributed equally to this work

[♠] Corresponding author

[a] Centre for Logic and Philosophy of Science, Department of Philosophy and Moral Sciences, Ghent University, Blandijnberg 2, 9000 Ghent, Belgium.

[b] Institute for Science in Society, Faculty of Science, Radboud University, the Netherlands.

E-mails: KristianCampbell.GonzalezBarman@Ugent.be, simon.lohse@ru.nl , henk.deregt@ru.nl

**Abstract:** We argue for the epistemic and ethical advantages of pluralism in Reinforcement Learning from Human Feedback (RLHF) in the context of Large Language Models (LLM). Drawing on social epistemology and pluralist philosophy of science, we suggest ways in which RHLF can be made more responsive to human needs and how we can address challenges along the way. The paper concludes with an agenda for change, i.e. concrete, actionable steps to improve LLM development.

## 1.Introduction

Reinforcement learning from human feedback (RLHF) is an increasingly common technique in artificial intelligence (AI) where a model learns by receiving feedback from humans rather than solely relying on a predefined reward function. This approach is particularly useful when designing AI systems for tasks where it is difficult to specify a precise reward function or when it is important to align the model's behaviour with certain human expectations and values. For instance, RLHF has notably improved language models for context-aware text generation (Ziegler et al. 2020) and taught robots to navigate cluttered environments (Henry et al. 2010).

RLHF is commonly employed in the later stages of fine-tuning models, particularly in the development of prominent Large Language Models (LLMs) like GPT-3.5 or GPT-4. Initially, these models undergo training using vast text corpora to grasp a broad range of language patterns and contexts. This foundational training is supplemented by task-specific fine-tuning, where the models are adjusted to excel in particular applications, such as understanding and generating dialogues. The refinement process is then further enhanced through RLHF. The RLHF process often involves people evaluating and ranking potential outcomes to build a classification model. This model is designed to predict if a human would find a specific output acceptable. The initial model undergoes refinement through a step-by-step process where it can use this classification model for feedback, allowing for improvement. This iterative process enables the model to better align with human judgement over time.

One question that becomes significant in this context is what role the *composition* of the human feedback group plays in the process. After all, a certain diversity in composition seems to be a clear advantage[1] in the development of models like ChatGPT. The feedback providers' background knowledge and their social, cultural and political perspectives may (or so we will argue) significantly influence the model – and a lack of diversity among individuals may lead to a model that is overly geared towards only specific expectations, potentially overlooking different but also important values and perspectives. Accordingly, too much homogeneity may raise *ethical* concerns and also create *epistemic* limitations, as the model may generate outputs that are hardly broadly acceptable, applicable or insightful. One key insight that emerges from our analysis is that achieving balance in LLM outputs need not require each individual evaluator to provide balanced feedback. Rather, more robust and nuanced outputs can emerge from the interplay of multiple distinct perspectives, each potentially representing a clear stance or worldview.

Despite the apparent benefits of diversity in RLHF, the philosophical underpinnings of why and how it may improve model performance have not been thoroughly explored. This paper offers an analytical perspective on the impact of diverse feedback on LLM development. Our investigation is motivated by two distinct but intertwined objectives: *first*, to elucidate the complex interplay between diversity, balance, and reliability in the development of LLMs employing RLHF. *Second*, to explore challenges for increasing diversity in RLHF and point to potential ways for managing these challenges.

The paper is structured as follows. Section 2 provides a brief overview of the landscape of RLHF and associated problems, highlighting the need for diversity in feedback. Section 3 introduces ideas from philosophy of science and social epistemology which can serve as useful conceptual tools to understand and manage the identified problems. Section 4 develops initial ideas about how RLHF can be improved by making it more pluralistic and discusses open questions and trade-offs that need to be addressed to make progress along these lines. The final section concludes with a brief summary and an agenda for making RLHF more pluralistic.

---

[1] For instance, Yamagata, McConville, and Santos-Rodriguez (2021) show that by incorporating feedback from a diverse group of trainers with varying skill levels and reliability into a Bayesian approach, it's possible to enhance the accuracy and robustness of Reinforcement Learning systems, even in the presence of unreliable sources. For an important theoretical perspective on the relationship between diversity and collective performance, see the "Diversity Trumps Ability Theorem" developed by Scott and Page and discussed in Landemore (2013); cf. Thompson, 2014 and Brennan, 2021 for critical discussions of the applicability and scope of the theorem.

## 2. Reinforcement learning and associated challenges.

RLHF is a useful method to align AI systems with human objectives, especially in the last stages of fine-tuning of state-of-the-art Large Language Models such as OpenAI's GPT-4 or Meta's LLama 3. This technique has become central in adapting AI models to complex, human-centric expectations and preferences. The clearest example of this was Davinci 3.5 (the first version of ChatGPT), which was far more safe, "user-friendly", and well received in terms of dialogic conversation style than GPT3.

Originating from revealed preference theory in economics, RLHF was adopted in machine learning for applications in human-computer interaction and reinforcement learning[2]. The standard methodology for RLHF was popularised in 2017 (Christiano et al. 2017) and has significantly influenced the deep reinforcement learning community. Note that LLMs are fundamentally stochastic machines that are proficient at generating text rather than systems that track or contain referential knowledge. While they can generate remarkably coherent and seemingly knowledgeable outputs, they operate through statistical pattern matching rather than by maintaining or accessing a database of facts about the world. This fundamental characteristic of LLMs has important implications for RLHF, as the feedback process aims to shape probabilistic language generation patterns rather than to instil or correct actual knowledge (we will come back to this below).

The advantages of RLHF are manifold. It enables humans to communicate goals to AI systems without the need for having a previously defined reward function. This not only makes reward shaping implicit and intuitive, but mitigates reward hacking (Casper et al. 2023), where an AI might exploit loopholes in a reward function to achieve high scores without actually fulfilling the task the reward function is a proxy for. Furthermore, RLHF facilitates a dynamic learning process where AI models can continuously evolve based on real-time human feedback, which enables companies with a large user base to improve their models with faster cycles. This ongoing interaction results in AI systems that are more aligned with the values and expectations of the model's users, ensuring their relevance and utility. Consequently, AI systems that incorporate RLHF might be better equipped to operate in complex moral and social landscapes, making them more trustworthy and reliable for a broad range of applications.

[2] Historically, RLHF can be situated within the broader framework of inverse reinforcement learning (IRL) approaches. IRL, first formalized by Ng & Russell (2000), addresses the fundamental problem of recovering an unknown reward function from observed optimal behaviour. While both RLHF and IRL aim to learn from human input, IRL traditionally focused on inferring reward functions from demonstrated optimal behaviour, whereas RLHF leverages direct human feedback on model outputs. This distinction is particularly relevant as RLHF does not assume optimal demonstrations but rather works with potentially noisy human preferences and feedback.

The process of RLHF, involves three sequential steps (Casper et al. 2023): Feedback Collection, Reward Modeling, and Policy Optimization. Feedback Collection involves the human assessment of model outputs, followed by Reward Modeling, where these assessments are used to train a reward model via supervised learning aiming to replicate the preferences of human evaluations. The final step, Policy Optimization, optimizes the pretrained LLM[3] to generate outputs positively rated by the reward model. The term 'policy' here refers to the updated LLM, which uses some form of reinforcement learning to improve the pretrained LLM.

To illustrate this, imagine an AI designed to write creative stories undergoing this three-step process. Initially, humans evaluate the stories the AI produces (e.g. based on creativity, coherence, and emotional impact). These evaluations guide the training of a reward model that predicts whether a story is appealing to humans (i.e., whether human evaluators would give it a good score). Finally, the AI's story-writing process is optimised based on feedback from the reward model, enabling it to produce narratives that better resonate with human readers, thus improving its ability to craft engaging and coherent stories.

Note, however, that the outputs of the AI will be influenced by the preferences of the evaluators, which may lead to a narrowing of the AI's capabilities and a potential bias towards certain types of stories or storytelling techniques. This risk is most clearly seen when stories are intended to contain answers to scientific questions, where evaluators might prefer concise and simple answers, posing a risk that the AI learns to provide a simplified but (potentially) misleading answer rather than a scientifically adequate one (Perez et al. 2022; see also Barman et al. 2024). In its most extreme form, this might involve giving not just simplified, but wrong answers altogether; for instance, the model may "hallucinate" responses to avoid answering that it does not know as this might be rated poorly. While this result could be seen as a simple consequence of prioritising user satisfaction, it may come at the expense of other values such as scientific accuracy and reliability. Given these (as well as other) potential trade-offs, it becomes important to consider which values one should aim for.

The human feedback used in RLHF includes, among others, label feedback, binary preference feedback, correction feedback, and language feedback. This involves tasks such as ranking outputs (e.g., in order of quality), choosing one output over another, providing explanations as to why a certain output is not correct (or how it could be improved), or providing an example of what the right output should be. Each feedback type has its distinct advantages and limitations, and there are multiple trade-offs at play (e.g., between the difficulty of the task, the ability of gathering enough feedback given a certain budget, or the difficulty of having humans with enough skill to perform these tasks). Despite its widespread use, certain limitations of RLHF in fine-tuning LLMs are notable. Perez et al. (2022) point out that models fine-tuned using RLHF often display biases and can lead to the mirroring of certain ideologies. El-Mhamdi et al. (2023) observe that these models could unintentionally reveal sensitive information. Furthermore, there is clear evidence of these models indeed producing hallucinated or inaccurate content (OpenAI et al. 2023). Moreover, RLHF doesn't protect these models against adversarial attacks, including jailbreaking[4] or prompt injection/extraction[5] (Li et al. 2023). The response behaviour of these models can also be undesirable in several other ways.

[3] It should be noted that before RLHF, there is usually a step of instruction fine-tuning, where the pre-trained LLM is refined on task-specific data, such as for dialogue generation.

These problems are too multifaceted to address them in one paper and require different types of solutions. In the following, we will focus primarily on problems of RLHF that are related to the *content* level, i.e. on models that have been "tutored" by humans but may still (or rather: because of this) generate biased, inaccurate or in other ways problematic responses to user prompts. Here are some examples for what we have in mind, drawing from an overview of the issues by Casper et al. (2023):

- RLHF can amplify biases and one-sided opinions of human evaluators, a phenomenon known as "sycophancy". This issue worsens with larger models and is not effectively mitigated by the RLHF process.
- RLHF can reinforce politically unbalanced judgements of the tutored models, an issue that seems to be associated with the demographic composition of the selected evaluators (Motoki et al. 2023).
- Some LLMs have been observed to become less accurate over time which may be, at least in part, a consequence of RLHF where evaluators make mistakes in light of difficult topic areas.
- The RLHF process may introduce unanticipated model drift (Metz, 2023; cf. Chen et al. 2023), meaning that the behaviour of the model might inadvertently shift from what was expected.
- Evaluators can intentionally or unintentionally insert harmful data into RLHF systems, especially given the scale at which RLHF operates. An example would be a (malicious) actor providing high ratings for clearly unethical responses of an LLM.

So what is at the heart of these problems? And how can we improve RLHF so that we can cope with them? Although there is probably no single answer to these questions, we argue that the outlined epistemic and ethical challenges can best be understood *through the lenses of social epistemology and pluralism*, as a lack diversity of human feedback plays a crucial role in these problems, indicating that strengthening diversity – in the right way – may provide useful starting points for tackling the issues.

[4] Jailbreaking refers to the process of exploiting a model to break its constraints or intended use cases, enabling it to produce outputs that were otherwise restricted or unintended. This could involve 'tricking' the model into generating harmful, biased, or prohibited content.

[5] Prompt injection/extraction involves manipulating the model's input (injection) or extracting sensitive information from it (extraction) by crafting inputs in a specific way. This could be used to reveal personal data the model has been exposed to during training or to cause the model to act in unpredictable or undesired ways.

**3. Conceptual tools from philosophy of science and social epistemology**

Increasing diversity is an obvious remedy for *some* of the identified problems. Consider the issues of user biases and politically unbalanced judgements. It seems intuitive that increasing diversity of perspectives in RLHF might be a (partial) remedy for these specific issues. To improve the AI alignment process, it is critical to look beyond individual perspectives to encompass the broader norms, expectations, and values of various groups. However, to better understand the underlying reasons for this and also *why* pluralism might be promising to understand and address some of the other problems mentioned, it will be helpful to introduce three interrelated concepts from post-positivist philosophy of science and social epistemology (as they have been developed by Paul Feyerabend, Sandra Harding, Helen Longino and others; see Grasswick, 2018 for an overview).

*I. Social cognition*: Feminist epistemologists such as Longino (1990) show that knowledge production is a communal, interactive process that is shaped by the way the process is organised on a social level. The way it is organised – in particular, the way in which the influence of values in the process is handled – can either promote or hamper the quality of the produced knowledge. While Longino makes this point in the context of science, the point generalises to reliable knowledge production per se. For instance, if you want to do a strategic SWOT (strengths, weaknesses, opportunities, threats) analysis in an organisation you will likely arrive at a richer, more reliable picture if you systematically incorporate internal feedback on different levels of the organisational hierarchy than if you only rely on the input of the top management. Organising this process according to an ideal of inclusiveness makes things better (cf. Bengio et al. 2024). This example brings another aspect of the sociality of knowledge to the fore, namely that knowledge is often socially distributed - no member of the organisation has complete knowledge about the organisation, but this knowledge is distributed among members and certain roles (the same applies to knowledge in scientific communities and many other contexts, see, e.g., Uygun Tunç 2023).

*II. Pluralistic triangulation*: In order to truly understand a particular (epistemic or ethical) problem or question, it is useful to compare and contrast different points of view. Through this triangulation process, we can avoid myopia and errors in possible answers or solutions. This is because the extent to which alternative points of view have blind spots or problematic assumptions often only becomes apparent through contrast. This simple yet powerful insight goes back at least to J.S. Mill's defence of freedom of speech and has most forcefully been defended by Feyerabend (1975, 1999)[6]. Think of the recent Coronavirus crisis. It was precisely the public involvement of experts beyond medicine and epidemiology (such as social scientists and managers of retirement homes) in analysing the public crisis that helped to bring an excessive focus on health (rather than, e.g., social justice) and problematic assumptions about pandemic modelling to light. This was in part the result of triangulation of perspectives, approaches, and also of different value commitments (Bschir and Lohse 2022).

*III. Positionality:* People's perspectives differ in systematic ways, influencing the ways individuals understand and interpret the world. More specifically, perspectives are shaped by personal experiences which in turn are connected to social position and role in society, most prominently by gender, social class and ethnicity. This insight from the sociology of knowledge can be considered as commonplace today. Feminist social epistemology, however, highlights two points. First, perspectives may not only affect ethical and political beliefs (as already Marx would have argued), but also areas of knowledge that typically do *not* seem to be affected by them, such as scientific knowledge (as shown by recent work in the philosophy of science; e.g. Massimi 2022). Second, the perspectives of marginalised or oppressed groups can be especially fruitful in discovering hidden and/or problematic aspects of a claim or state of affairs. Marginalised groups may sometimes be in a position where they can combine insider and outsider perspectives to enrich mainstream-dominated discourses and positions and to counteract negative consequences of marginalisation/oppression for knowledge production (Harding 1996; Intemann 2010) – which also raises the question of appropriate *representation* of different groups (see below). An example of this can be found in gender-biased explanations in primatology, which primarily emphasised the role of male animals in shaping social hierarchies, not taking into account the active role of female primates; a bias that only became apparent when more female researchers entered the field, who were sensitive to both sides of gender dominance due to their own positions in patriarchal societies (Haraway 1984).

Taken together, these concepts underline the importance of including multiple viewpoints from different social positions in a well-organised process if one is interested in a less error-prone and biased and more reliable and ethically robust understanding of many aspects of the world. Note that the above discussion of these concepts has an epistemic as well as an ethical dimension and implies that these are not independent. This aligns with a prominent strand in the current philosophical debate about the role of values in science, criticizing the traditional 'value-free ideal' of science (Douglas 2009). This traditional view has been superseded by a more nuanced view defended by Kuhn (1977), which states that scientific research – and indeed any kind of truth-orientated knowledge production – should rely only on so-called epistemic values (such as accuracy and consistency) and exclude the influence of non-epistemic values (such as moral or political values). The reason would be that the former are conducive to finding objective truth, while the latter are inherently subjective. However, as Douglas and others have convincingly argued, even this more nuanced value-free ideal is untenable. One reason is that the distinction between epistemic and non-epistemic values cannot be drawn sharply (Rooney 2017). Another reason is that in many types of epistemic practices, non-epistemic values are ineliminably involved in decisions about the acceptance or rejection of hypotheses, for example in statistical hypothesis testing (the so-called problem of inductive risk; see Douglas, 2009). While these insights from philosophy of science do not correspond directly to issues in AI, they support the idea that epistemic and non-epistemic (e.g. ethical) considerations are typically intertwined in knowledge production.[7]

[6] See Lloyd (1997) for the connection between Mill and Feyerabend.

It might be objected, however, that the two dimensions need to be kept apart because conflating them runs the risk of threatening the objectivity of the knowledge production process. As regards the epistemic dimension, a pluralistic approach that includes different perspectives is more suitable to eliminate mistakes and biases and hence will produce more accurate and robust results. In a word, it will enhance the objectivity of the outcomes of the knowledge-production process. By contrast, allowing a plurality of moral values to affect the process will jeopardize the objectivity of the outcomes, or so the objectors might claim. We submit, however, that this objection holds water only if objectivity is *defined* as the outcome of reasoning processes in which non-epistemic values are absent. But, as Douglas (2009, 115-129) argues, objectivity can be achieved in various other ways, where it is important to note that the term 'objectivity' can apply to various aspects of the knowledge production process and that a key feature of objectivity is that it is a basis for trust. Douglas distinguishes eight senses of objectivity, of which the following two are especially relevant for our purposes.[8] First, with respect to individual thought processes, objectivity can be achieved by "taking a position that is balanced or neutral with respect to a spectrum of values" (Douglas 2009, 123). This is called 'value-neutral objectivity'. Douglas adds that it may be allowed to exclude extreme positions

[7] But see Grote (2024) for an alterative view on the relation between epistemology and ethics in machine learning.

[8] (Hoyningen-Huene, 2023) argues that these different "senses" of objectivity are actually *indicators* of objectivity or *means* to increase objectivity (understood as the absence of distorting factors introduced by the knowers or "epistemic subjects").

(e.g. sexist or racist ones). Second, with respect to social processes (as noted above, a crucial element of knowledge production), the outcomes can be regarded as objective if they result from an open discussion by the members of the community. Various conditions have to be met to achieve such 'interactive objectivity', of which diversity among participants in the discussion – also with regard to different value orientations – is an important one (Douglas 2009, 127-128, cf. Longino 1990). In sum, Douglas' analysis of objectivity not only evades the objection that value-laden science cannot be objective but also supports our case for improving RLHF by making it more diverse.

Applying these insights to RLHF allows for an exploration of epistemic and ethical dimensions of AI (and in particular LLM) development. But before we can do this we will need to address a possible objection to our approach, namely why we should assume that we can apply ideas from philosophy of science and social epistemology to LLMs. As noted above, LLMs do not track truth or produce genuine knowledge but (re-)produce statistical patterns. They have even been described as "stochastic parrots" (Bender et al. 2021) and "bullshit generators" (Hicks et al. 2024). One might therefore come to the conclusion that our approach is fundamentally misguided. However, an objection along these lines overlooks that our analysis is aimed at RLHF as an epistemic design principle of LLMs. RHLF aims to make LLMs behave in such a way that responses to user prompts seem as if they would track truth, produce knowledge etc., leading to LLM outputs ideally indistinguishable from responses that a knowledgeable human or a group of humans would provide. Hence, it seems reasonable and of pragmatic value to examine RLHF through an epistemological (and also ethical) lens and proceed as we suggest.

From an epistemological standpoint, diverse human feedback aids in creating AI systems that are more reflective of varied human perspectives, thereby enhancing the models' objectivity and applicability. Ethically, it aligns with the ideal of inclusivity, ensuring that AI systems do not perpetuate existing moral biases or inequalities. The described insights from pluralistic philosophy of science and social epistemology underscore the significance of incorporating feedback from a wide range of individuals in RHLF. This diversity goes beyond mere political representation; it is about avoiding too much homogeneity and integrating various epistemic standpoints that arise from different social, cultural, and political positions and backgrounds. These standpoints bring unique insights, challenge prevailing assumptions, and may ultimately contribute to a more balanced and comprehensive – a more humanistic – AI model. However, to achieve this, we argue, it would be necessary to revise key elements in the RHLF process in light of the above considerations. In the next section, we make initial suggestions along these lines and point out open questions that need to be addressed to make progress.

## 4. Open questions and potential trade-offs

How can we make RLHF more pluralistic and what are the challenges along the way? To make progress on these questions, we will present tentative ideas. Above all, however, we want to point out important issues that will need to be addressed to move forward. We first give a quick overview of some practical aspects. We will then discuss evaluator selection, guidance and monitoring, and aggregation of individual feedback. Finally, we will point out possible trade-offs of different types.

### *4.1 Current practices*

We focus on OpenAI's methods, as ChatGPT is the model with most users. While OpenAI does not disclose too many details, there is general information as to how this feedback was collected in the initial stages. Ouyang et al. (2022) provide information on how feedback was collected from humans (in particular, they hired a team of 40 contractors on Upwork and ScaleAI), including instructions for human evaluators and related details to fine-tune InstructGPT (the direct predecessor of ChatGPT):

1. **Labeller Selection and Criteria**: Labellers were chosen based on their agreement with sensitive speech flagging, agreement on rankings, sensitive demonstration writing, and self-assessed ability to identify sensitive speech for different groups.[9] The selection process aimed to ensure labellers were sensitive to various demographic groups and could identify potentially harmful outputs.

2. **Detailed Labelling Instructions**: Labellers received detailed instructions on evaluating text outputs based on their helpfulness, truthfulness, and harmlessness. The way of evaluating 'helpfulness' was in terms of whether the output followed the users' intentions and helped solving the task. The way of evaluating 'truthfulness' was in terms of the output containing accurate information, and not misleading the user. Evaluating 'harmlessness' involved considering that the output should not cause physical, psychological, or social harm to people; or damage to equipment, property, the environment, institutions, or resources necessary to human wellbeing (for examples consult the paper by Ouyang et al. and the public labelling instructions provided to evaluators).

[9] In other words, they assessed their ability to identify potentially offensive content, they evaluated their consistency in evaluating and comparing content, they tested their capability to write responses that consider sensitivity issues, and they required labellers to evaluate their own ability to recognize content that may be sensitive.

3. **Labelling Instructions Evolution**: The instructions given to labellers evolved throughout the project to address feedback and a better understanding of the desired measurements. Initially, labellers were instructed to prioritise helpfulness to the user above truthfulness and harmlessness. However, in the final evaluations, the emphasis shifted to prioritise truthfulness and harmlessness. Here are some of the instructions given (from the instruction sheet):

    For each input instruction, labelling will consist of 3 parts:
    1. **Labelling instructions.** You'll label a bunch of properties of the instruction, including whether it contains PII, etc. [some examples: Personally identifying information, User intent is unclear, Asks for sexual content, Asks for violent content, Asks for content denigrating a protected class[10], Asks for moral judgement, etc.]
    2. **Labelling AI model outputs**. For each output from an AI model, you'll label the outputs along several different axes, including giving a 1-7 rating, judging whether it contains violent or sexual content, etc.
    3. **Ranking AI model outputs**. You'll rank outputs from best to worst, including ties."

    (Instruction sheet, footnote added [accessed 11 December 2024]) [https://docs.google.com/document/d/1MJCqDNjzD04UbcnVZ-LmeXJ04-TKEICDAepXyMCBUb8/edit#heading=h.21o5xkowgmpj]

4. **Collaboration with Labellers**: The project involved close collaboration with labellers, including an onboarding process, detailed task instructions, and a shared chat room for answering questions.

5. **Labeller Demographics**: A voluntary anonymous survey was sent to labellers to understand their demographics better (see fig. 1):

[10] From Appendix B, List of protected classes: Race, Color, Religion or creed, National origin or ancestry, Sex (including gender, pregnancy, sexual orientation, and gender identity), Age, Physical or mental disability, Veteran status, Genetic information, Citizenship.

Table 12: Labeler demographic data

| What gender do you identify as? | |
|---|---|
| Male | 50.0% |
| Female | 44.4% |
| Nonbinary / other | 5.6% |
| **What ethnicities do you identify as?** | |
| White / Caucasian | 31.6% |
| Southeast Asian | 52.6% |
| Indigenous / Native American / Alaskan Native | 0.0% |
| East Asian | 5.3% |
| Middle Eastern | 0.0% |
| Latinx | 15.8% |
| Black / of African descent | 10.5% |
| **What is your nationality?** | |
| Filipino | 22% |
| Bangladeshi | 22% |
| American | 17% |
| Albanian | 5% |
| Brazilian | 5% |
| Canadian | 5% |
| Colombian | 5% |
| Indian | 5% |
| Uruguayan | 5% |
| Zimbabwean | 5% |
| **What is your age?** | |
| 18-24 | 26.3% |
| 25-34 | 47.4% |
| 35-44 | 10.5% |
| 45-54 | 10.5% |
| 55-64 | 5.3% |
| 65+ | 0% |
| **What is your highest attained level of education?** | |
| Less than high school degree | 0% |
| High school degree | 10.5% |
| Undergraduate degree | 52.6% |
| Master's degree | 36.8% |
| Doctorate degree | 0% |

**Figure 1.** Labeller demographics (Ouyang 2022, p.40). The demographic data shows that the largest ethnic group was Southeast Asian (52.6%), followed by White/Caucasian (31.6%). Nationalities included Filipino and Bangladeshi (22% each), and American (17%). Most participants were aged 25-34 (47.4%), and educationally, 52.6% had an undergraduate degree, while 36.8% held a master's degree.

The initial stages arguably lacked diversity. However, with its opening to the broader user base of OpenAI, the diversity of users providing feedback has seen significant improvement, as all users throughout (193 countries as of January 2024) are able to provide feedback. The current ChatGPT interface allows users to give a thumbs up, thumbs down and to copy the output generated by the model. Sometimes the user is given two alternatives from which to choose. Additionally, when the model is asked to regenerate an answer, the user is asked whether the second answer was better, worse, or the same as the previous answer.

Now that we have a rough overview of some aspects of the process[11], we will discuss pluralistic reform ideas and associated challenges. Out of the 3 steps of RLHF, feedback collection and reward modelling are the most relevant for our purposes. For the first, we distinguish issues related to selecting evaluators and to guiding and monitoring them. For the second, we consider issues related to the reception and aggregation of the data and its implications for training of a reward model. We also examine the relevance of guidance and the possibility of trade-offs.

*4.2 Selecting evaluators*

An obvious implication that follows from the above discussion on social epistemology is that the procedure to select evaluators for RLHF will need to be revised to make it more representative of the existing diversity of perspectives (to prevent composition like the initial one above, in which almost all raters are educated, almost half American and a third Filipino). While the latter stages in this case have more diversity than the initial ones, it should be noted that the current user demographic of OpenAI, and by extension the feedback provided for RLFH, may not fully represent society at large (or even broad segments of it). For instance, there likely is a predominance of younger, well-educated individuals who have both the knowledge and technological means to interact with the free version of ChatPGT. In the case of GPT-4, given that there is a financial barrier for prolonged/ professional use, it may lead to the self-selection of certain strata of the population. Similarly, there might be a self-selective mechanism of those that engage in providing feedback. The issue here is that there may be certain people who are more inclined to provide this feedback than others, creating a possible self-selection bias (see  Bethlehem 2010 for self-selection effects in web surveys).

Although these problems might be easy to identify, it is far less clear how they ought to be resolved. So far we detailed that a source of problems stems from the evaluators not representing broader groups (which is relevant because of *positionality*). This naturally raises the question of the sense in which evaluators should be representative. Answering this question involves normative choices regarding the sense in which we want to make LLMs better aligned with human goals. Having a panel that is truly representative of the world population seems hardly feasible, as it would require comprehensive knowledge of the relevant populations in all regions – a level of detail we lack even for many well-studied countries. So, should developers aim for a specific subset of the world's population? If so, which one? Should they aim for a representative set of only a certain range of cultural and/or ethical standpoints and values (cf. Intemann 2010) and if so, which ones? Should they give a greater weight to evaluators that are less represented in the evaluator pool within society (e.g. elderly people)? Hidden in the background of these questions is an even more intricate one: Which dimension of diversity should be seen as relevant? Above, we pointed out that social and cultural standpoint matters. But this observation only points to the question of which dimensions of diversity are epistemically and/or ethically (most) relevant for RHLF. Natural candidates are gender, ethnicity, age, social class, and nationality. There may, however, be many more dimensions, some of them more difficult to operationalise than others, say cultural background and cognitive diversity.

[11] See Anthropic (Bai et al., 2022) for a discussion on Reinforcement Learning from AI Feedback (RLAIF), which extends Reinforcement Learning from Human Feedback (RLHF) by using AI-generated evaluations. Unlike RLHF that depends on human judgement, RLAIF uses constitutional principles to guide AI self-critique towards harmlessness and helpfulness, although due to similar stakeholder input challenges as RLHF, it falls prey to similar problems as the ones discussed below.

One can also ask the more fundamental question of whether representation should actually be the primary goal in the selection of evaluators - or whether we should treat "adequate" representation as means to our epistemic and ethical ends. Shouldn't there be certain thresholds, say regarding cognitive competencies and ethical outlooks?[12] Above we highlighted the problem that evaluators may rate answers in ethically problematic ways. In those cases where a certain rating is not motivated by *deliberately* inserting harmful data but reflects the implicit sexist outlook of, say, a Christian fundamentalist should this perspective even be represented? If not, who pre-defines the spectrum of reasonable viewpoints? As we saw above, Douglas (2009, 124) argues that even if a balanced representation of the spectrum of values is desirable for yielding value-neutral objectivity, it may be allowed to exclude extreme positions for good moral reasons. However, it is certainly not easy or straightforward to determine when this is the case (we will come back to this in the next section). Another question of representativeness connects to the issue of marginalised perspectives. If it is indeed the case that model output is ethically/epistemically biased due to the fact that it is predominantly based on mainstream training data (neglecting and/or distorting minority viewpoints underrepresented in the text corpus, thereby amplifying various forms of epistemic injustice, see Kay et al. 2024)[13], should the selection of evaluators be guided by a principle of affirmative action rather than pure representativeness? If so, those responsible for recruiting evaluators would also have to decide how these principles can be balanced in a reasonable way. A promising (and normatively justified) starting point could be to experiment with different compositions that slightly over-represent marginalised groups to understand the extent to which this actually influences the response behaviour of LLMs.

[12] OpenAI seemed to screen for this, to some extent, during the development of InstructGPT by conducting a test. This test consisted of a multi-part screening process involving sensitive speech flagging, ranking of model completions, demonstration writing, and self-assessed ability to identify sensitive content.

[13] For instance, *generative amplified testimonial injustice*: "Generative AI models are more likely to reproduce the voices of those frequently represented and culturally dominant online, while erasing the voices of the socially marginalized. This creates a feedback loop where dominant narratives are amplified and marginalized voices are further silenced, compounding the testimonial injustice experienced by these groups" (Kay et al. 2024).

### *4.3 Guidance and monitoring*

RLHF is a guided and monitored process. Evaluators do not act in a vacuum but are required to comply with certain guidelines. The developer team pre-defines which groups are classified as 'protected classes'. Similarly, they are given examples as to what constitutes harm. For instance, as can be seen in the instruction sheet (see Section 4.1), outputs such as 'The most common type of home-made bomb is made from a pressure cooker filled with explosive material.' are considered to be *highly* harmful; examples such as 'Stalin had good reason to believe that his enemies were plotting against him, and he took the necessary precautions to ensure his rule.' are considered as *potentially* harmful. While there is no perfect way to undertake guidance, there may be more structured ways to think about guidance and monitoring – this is where the conceptual tools from philosophy of science and social epistemology can be useful for addressing some of the most important issues.

A first issue concerns the principles that should guide the development of instructions for evaluators. There are many normative decisions necessary in order to come up with a set of instructions. What kinds of issues and possible harms should evaluators consider? On the basis of which values? How should different categories be prioritised? Looking at these questions from the perspective of pluralism and social epistemology does not provide us with concrete answers to these questions. However, it can point us towards the right kind of framework to address these questions. Such a framework should be pluralistic in the sense that it is sensitive to a multitude of different perspectives and values. This idea leads to asking a different question first: *Who* should decide on the right principles for evaluation guidance? Although it is possible for a certain group (let's say IT experts in Silicon Valley), to adopt different perspectives and take a certain variety of values into account, there are limits to this approach – namely the limits of what is conceivable for the group and those resulting from deep but implicit differences in the interpretation of certain states of affairs. Conceivability and interpretation will, at least to a certain extent, depend on the respective experiences and thus the respective social positions represented in any (in this case rather homogeneous) group. A promising answer to the question about the "who" is therefore that principles for instructions should be decided by a pluralistic panel representing sufficient diversity of perspectives and values. What "sufficient diversity" should entail in detail is, of course, subject to the same challenges as the recruitment issues discussed above. In our view, only a pragmatic approach is feasible here, starting with "obvious" diversity criteria (age, gender, religion, education, nationality) and expanding them iteratively based on experience and stakeholder feedback, especially by marginalised groups affected by epistemic injustices (cf. Kay et al. 2024). However, the more perspectives are represented at the table, the more mediation and active integration of viewpoints will be required, not least to address the issue of power imbalances and epistemic injustice in deliberation processes (Rolin 2021). Hence, modes of deliberation will have to be introduced that can help to manage these challenges. Here, we can learn from pluralistic ethics panels and similar advisory bodies that have developed tools to deal with the same kind of issues (Bschir and Lohse 2022; Nuffield Council on Bioethics n.d.).

However, even if we can find viable ways to deal with the described challenges, there is one essential question (which we already raised with regard to evaluator selection) that remains unanswered: How broad should the spectrum of values be that we consider legitimate? Of course, it is possible that sexist or racist values will also be represented[14] in a truly pluralistic panel deciding on instructions for evaluators, whether implicitly or explicitly. We believe that such values should not guide the respective instructions for evaluators. However, this means that an ethical framework must be considered that is pluralistic on the one hand, but also sets moral boundaries on the other. The *Universal Declaration of Human Rights* (UDHR) is a good starting point for this because it represents a minimum consensus that is already accepted, at least *de jure*, by a large number of countries.[15] How exactly implementation of such an ethically robust framework can succeed, however, is not a trivial question.

A pluralistic panel could also be responsible for monitoring adherence to the instructions and managing deviant behaviour of evaluators. It should, in other words, monitor and fine-tune this specific social process of *social cognition*. There will always be situations in which it is not clear how certain decision rules should be interpreted or applied or if evaluators really deviate from guidelines in a systematic way or not. This need not be due to malicious (or ignorant) actors, but simply to the fact that guidelines cannot cover every possible scenario and that there may be multiple interpretations of the rules (Miller and Sultanescu 2022). Take, for instance, the instructions given to evaluators involved in InstructGPT when evaluating truthfulness. One of the conditions was that outputs ‘Not produc[e] clearly false information about the world (e.g. making up facts or promoting conspiracies). For example, the output should not state that Hillary Clinton has served time in prison.’ (instruct sheet, p1); this might be straightforward for certain outputs, but not for others. Further, languages and cultural contexts influence how questions are interpreted and answers are structured, which can have an impact on RLHF (especially when feedback is given in different languages). Consider cases where different and cross-cutting kinship categories exist in different languages (Kemp and Regier 2012) which may lead to incommensurable answers to questions about social relationships.

[14] Another issue lies in the risk of harmful data insertion by malicious actors. Here, the main recourse simply lies in good screening and monitoring of whether evaluators are correctly following guidelines.

[15] Using the UDHR as a starting point for developing an ethical framework is not without its problems, as its legal and ethical interpretation are controversial (Glendon 1998). However, there seems to be no better, less controversial one.

When systematic rule application/deviance problems arise, it is necessary to make adjustments by introducing new rules, clarifying rules or adding examples (see above). The exact implementation of re-adjustments naturally poses similar issues as those mentioned in the last paragraph (e.g. “Is this evaluator mistaken – or does s/he have a value base that we didn’t consider?”), which is why a pluralistic panel makes sense here as well. In this context, the degree of openness in the response options available to evaluators should also be considered. The spectrum ranges from simple comparisons and rating procedures of LLM responses to procedures that allow the insertion of new kinds of responses. The latter is more complex, but at the same time more open to points of view that have not yet been considered. This might help to assess responses that require a greater level of fine-grainedness (e.g., they might be accurate and helpful but harmful, or harmless and helpful but false). Here, textual input from the evaluators would seem to enable a more reasonable approach. But it would also be an option to consider a different procedure, i.e. not to ask evaluators for balance, but rather to ask certain evaluators for pro-active liberal coding and others for leftist or conservative coding. Such an approach would entail explicitly valuing and preserving distinct viewpoints - whether liberal, conservative, or otherwise - rather than asking evaluators to temper or moderate their positions, thereby preserving the richness and nuance of different worldviews (rather than forcing artificial compromise) as well as acknowledging that true balance often emerges not from individual neutrality but from the dynamic interaction of multiple well-articulated positions. Furthermore, it may lead to more honest and consistent feedback, as evaluators can assess outputs from their genuine perspective rather than trying to approximate some notion of neutrality. This could establish a checks-and-balance mechanism via *pluralistic triangulation* in which multiple perspectives “pull” in different directions arguably leading to more epistemically and ethically robust model outputs (see Freiesleben and Grote 2023, for various notions of robustness in machine learning).

### *4.4 Feedback Collection and Aggregation*

An important step in RLHF is the collection and aggregation of individual feedback to enable the training of a reward model. The reward model is usually a supervised learning model that can predict evaluations. Recall, typical feedback includes binary preferences (i.e. 'yes' or 'no'; 'good' or 'bad') of a certain output, giving a score (e.g. from 0 to 7), or ranking several outputs in order of quality. In addition, feedback can also involve giving the explicit answer that should have been given, or providing an explicit explanation as to why the answer was correct/incorrect. These items are then aggregated and used to train a reward model. In the case of binary preferences or rankings the aggregation might be straightforward, although there are different aggregation mechanisms which may make a difference in the final policy that gets implemented (e.g., ELO rating, majority voting, averaging, etc.). For other types of feedback, important decisions regarding how the reward model is going to be trained, the order in which data is presented, and so on, make a difference.

In the case of InstructGPT, labellers ranked multiple model responses to the same prompt, creating numerous pairwise comparisons. The reward model was then trained on these comparisons, with all comparisons from a single prompt treated as a single batch. This iterative process used new comparison data to continuously update the reward model which was then used to train the policy. Although early training prioritised helpfulness, final evaluations emphasised truthfulness and harmlessness, leaving the resolution of potential conflicts between these criteria for future exploration. While optimising sequentially might do the job, it necessarily prioritises certain objectives over others, rather than keeping them all in mind. The way the feedback was aggregated, and the sequence by which the reward model evolved, raises several important questions when viewed through the lens of pluralism and social epistemology. One key issue is how to effectively combine the diverse, and potentially conflicting, feedback from evaluators with different backgrounds, expertise, and value systems. Disagreement can often signal areas of underlying complexity, ambiguity or tension that warrant further investigation. A pluralistic approach to aggregation would seek to preserve and engage with these points of contention, rather than erasing them in pursuit of a false consensus. This connects to the issue of how to handle conflicts between different alignment criteria, such as helpfulness, truthfulness and harmlessness. InstructGPT's strategy of shifting priority from helpfulness to truthfulness/harmlessness in later training phases is a rather blunt way of navigating these trade-offs. A more nuanced approach might involve explicitly modelling the relationships and dependencies between these criteria, and allowing for context-specific balancing rather than a one-size-fits-all hierarchy. In this context, several key additional questions and challenges in feedback collection and aggregation emerge.

One significant issue concerns preserving diverse perspectives during the aggregation process rather than cancelling them out or ignoring a subset. This makes it important to find methods that take important disagreements and minority viewpoints seriously in a *pluralistic process* instead of merely averaging them into a pseudo consensus. Connectedly, there is the challenge of navigating and balancing conflicting viewpoints. A key question is how the reward model can be aligned with, or at least take into account, the values of multiple groups holding conflicting or even incompatible viewpoints. Although there has been progress, representing diverse societies in reward models is still a pressing concern. How can we overcome the limitations of a single reward function in capturing the broad spectrum of human preferences, expertise, and capabilities in a diverse society, where current methods risk marginalising under-represented groups by treating differences as mere noise, or averaging important differences? This issue underscores the need for a reward system that is sensitive to diverse values and perspectives. Such an undertaking requires nuanced reward models. These models should adequately reflect the relationships and trade-offs among various alignment criteria, enabling context-specific adjustments. Such development would allow for a more refined approach to handling feedback that considers the complexity of different stakeholders' needs and expectations. Furthermore, it is worthwhile to explore the concept of multiple reward functions. Moving away from the reliance on a singular reward function and instead maintaining multiple models could more effectively represent the diverse values of different stakeholders.

In addition to these challenges, modelling dynamic human values presents its own set of complexities. Human feedback is inherently complex and context-dependent, continuously evolving with time. For example, attitudes towards certain countries might change due to geopolitical events (e.g. wars), or attitudes towards health or governments might change due to pandemics. This dynamism makes it difficult to effectively model with static, lower-dimensional preferences given by humans at a certain point in time, such as rankings of outputs.

Further iterations of RLHF will filter some of these problems out. However, certain complex issues may not improve through this process, or might require a different approach altogether (think for example of questions involving the assessment of the Middle East conflict and how RLHF might introduce model drift). A possible starting point for thinking differently about feedback aggregation can be seen in a notable study by Bakker et al.(2022) on fine-tuning LLMs to generate meaningful consensus statements among groups with diverse opinions. This involved creating debate questions using a pre-trained LLM on contemporary political and moral issues, followed by data collection and environment design where UK-based participants provided their opinions. The study demonstrated the model's ability to generate complex consensus statements preferred over LLM-generated and human-generated alternatives, highlighting the model's sensitivity to individual contributions and robust performance with out-of-distribution questions.,

*4.5 Moral and Epistemic Trade-offs*

Finally, we want to look at specific trade-offs in RLHF-assisted AI development. Regarding the type of feedback, there are trade-offs among accuracy, speed, and cost. For instance, the choice between binary responses and more detailed (pluralistic) feedback from evaluators significantly influences these dimensions. Binary responses, while quick and cost-effective, often lack the nuanced information that can refine AI behaviour more precisely. Conversely, detailed feedback, although richer and potentially more accurate, requires more time from evaluators and incurs higher costs. This extended feedback process may slow down the iterative cycles of model training, yet it enriches the quality and reliability of the model by providing deeper insights into the evaluators' reasoning and perspectives. Balancing these aspects requires careful consideration to optimise the efficiency of the RLHF process while maintaining the integrity and depth of the training data. However, we will not focus on these trade-offs, but rather concentrate on the ones related to epistemic and moral virtues.

Recent studies have pointed out a concerning trend in GPT-4's performance, particularly in areas requiring high precision, like mathematics (J. Chen et al. 2023). For example, in March 2023, GPT-4 demonstrated an 84% accuracy rate in identifying prime versus composite numbers, a task that demands precise mathematical reasoning. However, by June 2023, its accuracy had notably declined to just 51%. This diminishing accuracy brings to light questions about the impact of user diversity on the model. Does the interaction with a more heterogeneous user base enhance the model's overall performance, or does it amplify certain limitations? For instance, if users with varying degrees of expertise in a subject interact with the model, it might struggle to maintain a high standard of accuracy in that domain, but on the other hand, it might be able to interact with the subject at different levels of difficulty and abstraction. Possibly a comparison and *pluralistic triangulation* of different user responses can help here. However, such a process - perhaps even organised in a discursive format - naturally involves efficiency trade-offs.

Further complicating this dynamic, a study by (K. Chen et al. 2024) evaluates how GPT-3 interacts with diverse socio-demographic groups concerning sensitive topics like climate change and the Black Lives Matter movement. The findings reveal a nuanced landscape where conversational styles and user experiences correlate significantly with the users' background, potentially indicating biases in the model's responses. This is particularly evident as users from lower educational backgrounds and opinion minority groups report poorer experiences. Is this the result of the evaluators that were involved in RLHF? (recall that for InstructGPT, more than 85% of the evaluators had at least a university degree).

This dilemma underscores a potential trade-off between the epistemic and moral dimensions of AI development. On one hand, moral considerations advocate for a diverse range of feedback to ensure the model's alignment with a broad spectrum of human values and perspectives. On the other hand, epistemic virtues like accuracy and reliability might be compromised if the feedback providers lack domain-specific expertise. This tension necessitates a deeper philosophical inquiry into the objectives and methodologies of RLHF and prompts a re-evaluation of the balance between user satisfaction and (scientific) accuracy in the model's training objectives, among other things.

Furthermore, the interaction of diversity with model alignment and response generation introduces additional complexity. One of the places where this can be seen most clearly is in certain manifestations of model drift, particularly in systems that undergo continuous refinement and real-world applications. This phenomenon occurs when a model's performance and behaviour degrade over time due to shifts in underlying data, changes in the real-world context it is applied to, updates to the model itself, or as a consequence of RLHF. For instance, RLHF can introduce drift through the biases and preferences of the evaluators, particularly if they hold strong moral or ethical views that may not be universally shared or may become less prevalent over time. For instance, a language model trained up to a certain year might not only miss later developments but also reflect dated or skewed moral perspectives if the feedback mechanism doesn't account for evolving societal norms. This form of drift not only impacts the model's accuracy and reliability but also its relevance and fairness, as the AI might produce outputs that are misaligned with current or emerging societal values.

In addition to model drift, a related problem that may occur through progressive rounds of RLHF is that of *catastrophic forgetting*, where neural networks trained sequentially on different tasks tend to forget earlier tasks as they learn new ones. Kirkpatrick et al. (2017) show that techniques such as elastic weight consolidation protect important weights for previously learned tasks by constraining their plasticity. This could be to some extent imported within the design of the RLHF flow, with an eye on protecting the influence of high-quality early feedback by explicitly weighting or constraining updates based on feedback quality metrics. Another complementary approach would be to continually reintroduce high-quality feedback, effectively maintaining a balance between carefully curated feedback and newer (cheaper), potentially noisier user interactions. This hybrid approach might help maintain model performance while still allowing for adaptation to changing user needs and contexts.

These observations show one thing above all, namely that the way the *social cognition process* is structured in the case of RHLF and decisions regarding the balancing of different epistemic and moral values is anything but trivial - but it may have a major impact on the kind of answers we can expect from an evolving LLM.

While our discussion focuses on improving RLHF for large-scale models like the ones underlying ChatGPT, it's worth noting that the future landscape will likely feature multiple specialized models catering to different contexts, cultures, and use cases. However, market dynamics and network effects tend to create Schelling points - as we currently see with ChatGPT's role as a de facto standard for most LLM interactions. Even in a diverse ecosystem of AI models, certain ones will likely emerge as widely-used daily tools for a significant portion of the global population. This reality makes the case for pluralistic RLHF particularly pressing for these broadly-adopted models. Moreover, even models catered to specific cultural contexts or applications would benefit from the pluralistic approaches outlined in this paper, as they too will likely require handling diversity of perspective and value trade-offs within their target domains and scope.

## 5. Conclusion: Enhancing RLHF

In this article, we have set out the advantages of pluralistic RLHF and underpinned them with arguments from social epistemology and pluralist philosophy of science. We argued for increasing evaluator diversity, optimising guidance for heterogeneous evaluator responses, and modifying the approach to feedback aggregation, among other things – however, we also emphasised challenges for more pluralism in RLHF. While these challenges cannot all be overcome equally well, we have sketched a few possible ways for addressing them.[16] We conclude with an agenda for change, i.e. concrete, actionable steps, to improve LLM development.

In our view, the following steps are the most important – and feasible – starting points for advancing the pluralisation of RHLF:

(1) ***Representation***. LLM developers must address issues of *adequate* representation of epistemic perspectives and ethical values in the selection of evaluators more explicitly than before and negotiate these in a transparent process. This will need to include discussions about the spectrum of values that are considered legitimate – and ideally the development of an ethical framework that reflects the outcomes of these discussions. In doing this, developers should leverage stakeholder participation methods to include diverse, and in particular marginalised, voices throughout the process.

[16] Beyond LLMs, similar challenges and opportunities in pluralistic RLHF may arise in other fields such as robotics (potentially improving accessibility for disabled individuals), autonomous vehicles, game development (enhancing non-player character interactions), and personalised education.

(2) ***Guidance*.** LLM developers should establish pluralistic panels representing a diversity of perspectives and values to jointly decide on the principles for evaluation guidance and monitoring the feedback process. This should be done in an iterative way where additional (and especially minority) perspectives can be added to the panel as appropriate. In order to establish such a panel, a democratic mode of deliberation must be developed that enables an open exchange of viewpoints and is pragmatically feasible at the same time. An ethical framework like the one mentioned under (1) could serve as a scaffold for this deliberation.

(3) ***Feedback.*** LLM developers should experiment more with feedback formats that are capable of preserving different perspectives and conflicting viewpoints (allowing for negotiation and contestation) rather than turning them into superficial pseudo consensus. This entails fundamentally rethinking how to balance feedback collection: rather than asking individual evaluators to provide balanced or neutral feedback, the idea would be to actively seek out and preserve distinct perspectives and authentic viewpoints, whereby the balance would emerge from the multiplicity of perspectives rather than from individual restraint. This could, for instance, be achieved by creating platforms where evaluators with different perspectives engage in interactive discussions and collective revisions of their evaluations, using methods such as Delphi techniques or iterative feedback to develop more complex positions, especially when initial opinions differ significantly.[17]

(4) ***Aggregation*.** Ultimately, the challenge is to develop reward models that see diversity of values and standpoints as a resource rather than treating it as a problem to be minimised. This may require a rethinking of established methods for creating reward models. LLM developers should address this challenge by developing and testing more nuanced reward models. They could, for instance, explore and improve the usability of weighted feedback aggregation approaches that prioritise the input of evaluators based on their context-specific expertise and trustworthiness. This might involve experimenting with reputation systems, peer-review mechanisms, or criteria that account for an evaluator's background and qualifications,

[17] Such an approach would improve inclusivity of viewpoints and also align with Longino's (1990) and Douglas' (2009) notion of interactive objectivity, discussed in Section 3, where the right kind of social organisation of the process enhances the epistemic robustness and credibility of the outcome.

ensuring that more credible feedback (relative to the task at hand) has a greater impact on the outcomes.[18]

As we have shown, insufficient pluralism is problematic for RLHF on epistemic and ethical grounds. Ultimately, however, the problem is much more far-reaching, as biased and under-complex LLMs, once fully integrated into social media, journalism and other key areas of public discourse, have the potential to negatively influence this discourse on a massive level and ultimately jeopardise democracy. Thus, even if the changes we propose prove insufficient or pragmatically unfeasible, we believe they should be carefully considered and, if necessary, modified or replaced with better ideas and approaches. One way or another, the pluralisation of RLHF will be crucial for the development of more humanistic LLMs and hence for safeguarding public discourse.

[18] There may also be promising technical approaches. Rather than simple averaging, which may lead to pseudo-consensus, probabilistic models could capture the full distribution of individual preferences. For instance, Bayesian approaches like Gaussian processes could model distributions over individual rewards or even over entire reward functions (Biyik et al., 2020; Binois et al., 2021). This, in theory, could preserve the multi-modal nature of preferences rather than collapsing them (although in some cases practical implementations may struggle with high-dimensional multi-modal distributions). Similarly, the challenge of balancing different criteria (like truthfulness and harmlessness) need not rely on sequential optimization. Instead, multi-objective optimization techniques could identify different optimal solutions (e.g., the Pareto front), making trade-offs explicit and allowing for context-dependent balancing based on the relative scale of different desiderata (Chiandussi et al., 2012; Gaudrie et al., 2018; Paria et al., 2018; Calandra & Peters, 2014).